\documentclass[12pt]{article}
\usepackage{graphicx}

\newcommand\pubnumber{SLAC--PUB--14320}
\newcommand\pubdate{December, 2010}

\def\SLAC{SLAC, 
    Stanford University, Menlo Park, CA 94025 USA}
\def\doeack{\footnote{Work supported by the US Department of Energy,
                     contract DE--AC02--76SF00515.}}

\def\Title#1{\begin{center} {\Large #1 } \end{center}}
\def\Author#1{\begin{center}{ \sc #1} \end{center}}
\def\Address#1{\begin{center}{ \it #1} \end{center}}

\def\submit#1{\begin{center}Submitted to {\sl #1} \end{center}}
\newcommand\pubblock{\rightline{\begin{tabular}{l} \pubnumber\\
         \pubdate \end{tabular}}}
\newenvironment{Abstract}{\begin{quotation} \begin{center}
                       ABSTRACT
     \end{center}\bigskip  }{\end{quotation}}

\def\submit#1{\begin{center}Submitted to {\sl #1} \end{center}}
\def\Acknowledgements{\bigskip  \bigskip \begin{center} \begin{large}
             \bf ACKNOWLEDGEMENTS \end{large}\end{center}}
\def\beq{\begin{equation}}
\def\eeq#1{\label{#1}\end{equation}}
\def\eeqn{\end{equation}}

\newenvironment{Eqnarray}%
   {\arraycolsep 0.14em\begin{eqnarray}}{\end{eqnarray}}
\def\beqa{\begin{Eqnarray}}
\def\eeqa#1{\label{#1}\end{Eqnarray}}
\def\eeqan{\end{Eqnarray}}
\def\CR{\nonumber \\ }

\def\leqn#1{(\ref{#1})}

\let\bar=\overbar

\def\lsim{\mathrel{\raise.3ex\hbox{$<$\kern-.75em\lower1ex\hbox{$\sim$}}}}
\def\gsim{\mathrel{\raise.3ex\hbox{$>$\kern-.75em\lower1ex\hbox{$\sim$}}}}

\def\L{{\cal L}}

\def\half{\frac{1}{2}}

\def\del{\partial}
\def\Dslash{\not{\hbox{\kern-4pt $D$}}}
\def\dslash{\not{\hbox{\kern-2pt $\del$}}}
\def\pslash{\not{\hbox{\kern-2pt $p$}}}
\def\qslash{\not{\hbox{\kern-2pt $q$}}}
\def\kslash{\not{\hbox{\kern-2pt $k$}}}
\def\oneslash{\not{\hbox{\kern-2pt $1$}}}
\def\twoslash{\not{\hbox{\kern-2pt $2$}}}
\def\threeslash{\not{\hbox{\kern-2pt $3$}}}
\def\fourslash{\not{\hbox{\kern-2pt $4$}}}
\def\aslash{\not{\hbox{\kern-2pt $a$}}}
\def\rslash{\not{\hbox{\kern-2pt $r$}}}
\def\Jslash{\not{\hbox{\kern-2pt $J$}}}

\def\msb{{\bar{\scriptsize M \kern -1pt S}}}

\def\drb{{\bar{\scriptsize D \kern -1pt R}}}

\def\spa#1#2{ \langle #1 #2 \rangle }
\def\spb#1#2{ [ #1 #2 ] }
\def\apb#1 {  \langle #1 ] }
\def\bpa#1{  [ #1 \rangle }

\makeatletter
\def\section{\@startsection{section}{0}{\z@}{5.5ex plus .5ex minus
 1.5ex}{2.3ex plus .2ex}{\large\bf}}
\def\subsection{\@startsection{subsection}{1}{\z@}{3.5ex plus .5ex minus
 1.5ex}{1.3ex plus .2ex}{\normalsize\bf}}
\def\subsubsection{\@startsection{subsubsection}{2}{\z@}{-3.5ex plus
-1ex minus  -.2ex}{2.3ex plus .2ex}{\normalsize\sl}}

\renewcommand{\@makecaption}[2]{%
   \vskip 10pt
   \setbox\@tempboxa\hbox{\small #1: #2}
   \ifdim \wd\@tempboxa >\hsize     % IF longer than one line:
       \small #1: #2\par          %   THEN set as ordinary paragraph.
     \else                        %   ELSE  center.
       \hbox to\hsize{\hfil\box\@tempboxa\hfil}
   \fi}

 \def\citenum#1{{\def\@cite##1##2{##1}\cite{#1}}}
\newcount\@tempcntc
\def\@citex[#1]#2{\if@filesw\immediate\write\@auxout{\string\citation{#2}}\fi
  \@tempcnta\z@\@tempcntb\m@ne\def\@citea{}\@cite{\@for\@citeb:=#2\do
    {\@ifundefined
       {b@\@citeb}{\@citeo\@tempcntb\m@ne\@citea\def\@citea{,}{\bf ?}\@warning
       {Citation `\@citeb' on page \thepage \space undefined}}%
    {\setbox\z@\hbox{\global\@tempcntc0\csname b@\@citeb\endcsname\relax}%
     \ifnum\@tempcntc=\z@ \@citeo\@tempcntb\m@ne
       \@citea\def\@citea{,}\hbox{\csname b@\@citeb\endcsname}%
     \else
      \advance\@tempcntb\@ne
      \ifnum\@tempcntb=\@tempcntc
      \else\advance\@tempcntb\m@ne\@citeo
      \@tempcnta\@tempcntc\@tempcntb\@tempcntc\fi\fi}}\@citeo}{#1}}
\def\@citeo{\ifnum\@tempcnta>\@tempcntb\else\@citea\def\@citea{,}%
  \ifnum\@tempcnta=\@tempcntb\the\@tempcnta\else
  {\advance\@tempcnta\@ne\ifnum\@tempcnta=\@tempcntb \else\def\@citea{--}\fi
    \advance\@tempcnta\m@ne\the\@tempcnta\@citea\the\@tempcntb}\fi\fi}
\makeatother

\def\spa#1#2{\langle #1 #2 \rangle}
\def\spb#1#2{[ #1 #2 ]}
\def\apb#1#2#3{\langle #1  #2  #3 ]}
\def\bpa#1#2#3{[ #1   #2  #3 \rangle}

\def\Res{\mbox{Res}}
\def\L{{\cal L}}

\begin{document}
\begin{titlepage}
\pubblock

\vfill
\Title{Top Quark Amplitudes with an Anomalous Magnetic Moment}
\vfill
\Author{Andrew J. Larkoski and Michael E. Peskin\doeack}
\Address{\SLAC}
\vfill
\begin{Abstract}

The anomalous magnetic moment of the top quark may be measured during the 
first run of the LHC at 7 TeV.  For these measurements, it will be useful
to have available
 tree amplitudes with $t\bar t$ and arbitrarily many photons and 
gluons,
including both QED and color anomalous magnetic moments.  In this paper, 
we present a method for computing these amplitudes using the 
Britto-Cachazo-Feng-Witten (BCFW)
recursion formula.  Because we deal with an effective theory with 
higher-dimension couplings, there are roadblocks to a direct computation
with the BCFW method.  We evade these by using an auxiliary scalar 
theory to compute a subset of the amplitudes.  

\end{Abstract}
\vfill
\submit{Physical Review D}
\vfill
\end{titlepage}
\def\thefootnote{\fnsymbol{footnote}}
\setcounter{footnote}{1}
\tableofcontents
\newpage
\setcounter{page}{1}

\section{Introduction}

The first run of the Large Hadron Collider (LHC)
 at 7 TeV promises to yield a wealth of data 
and could lead to hints at physics beyond the Standard Model.  
While we do not know what questions the LHC will answer 
in regards to electroweak symmetry breaking, supersymmetry, 
dark matter or other new physics, we can be sure that during 
the first run our knowledge of the Standard Model particles will increase. 
 In particular, the number of top quarks that will be produced at the LHC 
will be comparable to that produced so far at the  Fermilab Tevatron,
and a much greater sample will be produced at large $t\bar t$ masses.
This will give us an opportunity to probe for interactions of the 
top quark that might indicate its composite structure or coupling to new 
forces.  

One aspect of this study will be the search for anomalous magnetic moment 
couplings of the top quark.  The consequences of anomalous magnetic moments of
the top quark have been considered previously, beginning with the work
of Atwood, Kagan, and Rizzo~\cite{AKR} and Haberl, Nachtmann, and 
Wilch~\cite{HNW}. These authors analyzed top quark pair production; they
computed
the effect of the color anomalous magnetic moment on the 
total cross section and distributions of the top 
quarks for this process. 
At the high energies available at the LHC, however, one 
should also 
consider the effect of radiation of additional gluons.  It would be useful
to have a calculational method that could produce arbitrarily complicated
tree amplitudes of this type.

In this paper, we will discuss a straightforward method for computing
$t \bar t + n g + m \gamma$ tree amplitudes of arbitrary complexity.
In principle, these amplitudes can be computed from Feynman diagrams. 
However, the multiple vertices and the complexity of gluon interactions
make this a challenge.  Already at the level of $t\bar t + 4 g$ processes,
corresponding to $t\bar t$ production with 2 gluons radiated, 
there are over 100 Feynman diagrams.  This number increases greater than 
factorially
with the number of gluons.  A better solution would be to compute the 
amplitudes
recursively, using either the Berends-Giele approach~\cite{BG} or the 
more recently proposed on-shell recursion formula of Britto, Cachazo, Feng,
and Witten (BCFW)~\cite{BCFW}.  Some time ago, Schwinn and Weinzierl
 developed a
formalism for massive quarks that uses the BCFW method and is
computationally very effective for QCD tree amplitudes~\cite{Schwinn}.

However, the Schwinn-Weinzierl scheme does not generalize directly to include 
the anomalous magnetic moment couplings. The BCFW method requires good
behavior of amplitudes as some external momenta are taken to infinity.
Thus it is nontrivial to apply this method to effective Lagrangians that
involve higher-dimension interactions.  Indeed, we find that direct 
application of the BCFW method is stymied by the additional momentum factor
in the anomalous magnetic moment vertex.

Fortunately, there is a way around this difficulty.  We find that 
those amplitudes that cannot be computed by direct application of BCFW
can be computed using an auxiliary theory of a scalar particle that 
carries the spin internally.  Combining the results, we produce a compact
recursive method.    This method introduces what we consider a promising
approach to the application of on-shell recursion to general effective
Lagrangians with higher-dimensional interactions. 

The outline of this paper is as follows:  In Section 2, we will present
our notation and review some aspects of BCFW computation.  In Section 3,
we will analyze the use of BCFW recursion for fermions with anomalous 
magnetic moment couplings.  In Section 4, we will present a useful 
rewriting of this theory as an auxiliary scalar theory.  In Section 5, we 
will present some explicit calculations that check the relation of this
scalar theory to the original fermion theory.  In Section 6, we will 
present our conclusions and compare our approach to other work on the 
treatment of higher-dimension interactions by on-shell methods.

\section{Review of Spinor Helicity and BCFW Recursion}\label{rev}

The goal of this paper will be to present a method for tree-level 
calculations in the theory
\beq
 \L = \bar \Psi \biggl[  i \Dslash - m  + {g a \over 4m} 
              \Sigma_{\mu\nu}F^{\mu\nu a} t^a \biggr] \Psi \ ,
\eeq{basicL}
 where $g$ is the QCD coupling, $a$ is a fixed constant color
anomalous magnetic
moment, $F^{\mu\nu a}$ is the QCD field strength, 
and $\Sigma_{\mu\nu} = i [\gamma_\mu,\gamma_\nu]/2$. The same
method will generalize readily if the theory also includes a
 QED anomalous magnetic moment term
\beq
 \delta \L = + \bar \Psi \biggl[ {2\over 3}
       { e a_{\mbox{\scriptsize QED}}\over 4m }
         \Sigma_{\mu\nu}F^{\mu\nu}\biggr] \Psi \ .
\eeq{QEDanom}

Throughout this paper, we will use the spinor helicity notation, as
 reviewed pedagogically
 in \cite{spinors}.  Instead of using 4-vectors, we will 
use as 
 fundamental objects the spinor products
\beq
\spa ij = \bar{u}_-(i)u_+(j)\qquad \spb ij=\bar{u}_+(i)u_-(j) \ .
\eeq{spinordef}
associated with lightlike vectors $p_i$, $p_j$.
These objects are antisymmetric and obey
\beq
|\spa ij|^2=|\spb ij|^2= (p_i + p_j)^2 = s_{ij} \ .
\eeq{spinorsq}
The spinor completeness relation is written in this language as
\beq
p\rangle[p+p]\langle p=\pslash \ .
\eeq{spinorcomp}
As an example, the polarization vectors of a gauge boson can be written as
\beq
\epsilon_+^\mu(k)={\langle r \gamma^\mu k] \over \sqrt{2}\spa rk} 
\qquad \epsilon_-^\mu(k)=- {[ s \gamma^\mu i\rangle \over \sqrt{2}\spb sk}\ ,
\eeq{spinorpol}
using auxiliary reference spinors $r$, $s$.
The spinors $r$, $s$ are arbitrary, corresponding to the
 gauge freedom of the boson.  

The spinor helicity formalism has been extended for use with massive 
fermions by Schwinn and Weinzierl (SW)~\cite{Schwinn}.  
For a massless fermion, the helicity states are 
physically distinct and Lorentz-invariant.  For massive fermions, there
is no unambiguous specification of spin state.  In the formalism of SW,
a lightlike reference vector $r$ 
is used to specify the spin basis to be used.  Starting with the massive
4-vector $p$, one defines
a lightlike 4-vector $p^\flat$  by 
\beq
\pslash^\flat= p^\flat\rangle [p^\flat = \pslash 
      - {m^2 \over \apb{r}{p}{r}} \rslash \ .
\eeq{pflatdef}
Then the $u_+(p)$, $u_-(p)$ spinors for a massive fermion are
\beq
u_+(p)={(p+m)r\rangle \over \spa{p^\flat}{r}} \qquad 
u_-(p)={(p+m)r] \over \spb{p^\flat}{r}}  \ .
\eeq{spinorferm}
It is straightforward to check that these spinors satisfy the required
completeness relation. 

We will express the values of fermion amplitudes by taking all 
fermions and antifermions to be outgoing.  With this prescription,
outgoing fermions are described by spinors $\bar u(p)$ given by
\beq
  \mbox{for}\ q_R\ : \ {\langle r (p+m)\over \spa r{p^\flat} } 
 \ , \qquad  \mbox{for}\ q_L\ : \ { [r (p+m) \over [rp^\flat]} \ ,
\eeq{qspinors}
Outgoing antifermions are described by spinors $v(p)$ given by 
\beq
  \mbox{for}\ \bar q_R\ : \ { (p-m)r \rangle\over \spa {p^\flat} r} 
 \ , \qquad  \mbox{for}\ \bar q_L\ : \ { (p-m)r] \over [p^\flat r]} \ .
\eeq{qbarspinors}

 To study the effects of top quark polarization, it
is useful to be able to compute massive fermion amplitudes for an 
arbitrary choice of the reference vector $r$ for each fermion.  We will 
try to retain that freedom in our analysis.

BCFW \cite{BCFW} 
proposed a method for computing QCD amplitudes based on the idea
of deforming the external momenta by a complex parameter $z$  such that
total momentum remains conserved and 
 all particles remain on-shell.  The explicit deformation they proposed
chooses two particles $i$, $j$ and modifies their momenta according to
\beq
p_i\to p_i - zq \ , \qquad
p_j \to p_j+ zq \ .
\eeq{pshift}
To keep particles $i$ and $j$ on-shell, $q$ must be light-like and 
satisfy $q\cdot p_i=q\cdot p_j=0$.  For massless $i$ and $j$ 
this can be expressed as a deformation of the individual spinor components:
\beqa
i\rangle\to i\rangle- z\: j\rangle \ ,\qquad i] \to i]\ ,\CR
j\rangle \to j\rangle\ ,\qquad j]\to j]+z\: i] \ . 
\eeqa{spinorshift}

At tree level, the deformed amplitude ${\cal A}(z)$ has 
only simple poles in $z$ from Feynman propagators going 
on-shell.  BCFW then considered the object
\beq
\oint {dz \over z} {\cal A}(z) \ ,
\eeq{BCFWint}
where the contour encircles $z=0$ and is taken to $\infty$.  
If ${\cal A}(z) \to 0$ as $z\to \infty$, 
the integral receives no contribution from the contour at 
$\infty$ and the integral vanishes.  By Cauchy's theorem, 
this is the sum of residues of poles in the contour.  Then,
\beq
{\cal A}(0)= - \sum \Res\;{\cal A}(z) \ .
\eeq{preBCFW}
The quantity on the left-hand side of \leqn{preBCFW} is the original 
amplitude to be evaluated.
The residues on the right occur when the 
deformed momentum that flows through a propagator 
goes on-shell.  This relates lower point on-shell 
amplitudes to the amplitude of interest.  BCFW thus obtain a 
recursion formula that allows the original amplitude ${\cal A}$ to 
be computed in terms of lower-point amplitudes.

More explicitly, the recursion relation is
\beq
i{\cal A}=\sum_{\mbox{\scriptsize cuts}} 
i{\cal A}_{\mbox{\scriptsize L}}(\hat{i})\ 
{i \over P_{\mbox{\scriptsize L}}^2}\ 
i{\cal A}_{\mbox{\scriptsize R}}(\hat{j}) \ .
\eeq{BCFWdef}
The sum runs over cuts through a single propagator that divide the amplitude
into two parts, with the external leg $i$ in the left-hand amplitude
${\cal A}_{\mbox{\scriptsize L}}(\hat{i})$ and 
the external leg $j$ in the right-hand
amplitude
${\cal A}_{\mbox{\scriptsize R}}(\hat{j})$.   These amplitudes  are computed 
with all momenta on-shell and with $i$ and $j$ set 
to their shifted values.  The identity requires good  
large $z$ behavior of the amplitude 
${\cal A}(z)$.  If this amplitude does not tend to zero as $z\to \infty$, 
extra terms appear from the contour at $\infty$ that invalidate the 
simple recursion.

\section{Large $z$ behavior}\label{fermBCFW}

Since the BCFW recursion formula depends on good behavior of the shifted
amplitude as $z \to \infty$, there is a danger that the BCFW method will
not be valid for effective theories that contain non-renormalizable 
operators.  In this section, we will show that this is a problem for the 
model \leqn{basicL}.  Specifically, we will show that tree amplitudes 
in the theory \leqn{basicL} can be computed in terms of amplitudes with 
all $+$ or $-$ gluon helicities.  However, this still leaves a gap that
needs to be filled before all amplitudes can be computed from simple
components.

In our analysis of the theory \leqn{basicL}, we will only consider 
shifts of gluon momenta.  In \cite{Schwinn}, SW give a prescription for 
shifting the momenta of external massive fermions.  However, this analysis
works only for specific choices of the reference vector $r$ in \leqn{pflatdef},
while we would like to maintain the freedom to work with an arbitrary
choice of $r$. 

 An arbitrary shift on gluons will, according to 
\leqn{spinorshift}, involve an external momentum $i$ with its angle
bracket shifted and an external momentum $j$ with its square bracket 
shifted. There are four possible helicity combinations of the  $ij$ to 
consider: $++$, $+-$, $--$ and $-+$.  For standard QCD with $a = 0$,
the first three shifts give good $z\to \infty$ behavior while the last
case $-+$ does not allow BCFW recursion.  Still, for any two gluons, there
is an allowed shift, and so any $q\bar q + n g$ amplitude can be 
reduced to 3-point functions by BCFW recursion.

Now consider adding to the theory the anomalous magnetic moment vertex.
If the shift momentum $zq$ flows into the quark line through this 
vertex, the vertex is proportional to $z$ at large $z$.  A fermion 
propagator carrying the shift momentum behaves as $z^0$, and all other 
fermion vertices---including the magnetic moment vertex with an ${\cal O}(1)$
external momentum---scale as $z^0$.  If the gluon from the magnetic 
moment vertex is connected to external gluon lines through a tree of gluons,
each propagator in this tree carrying the shift momentum scales as $z^{-1}$
and each vertex is at worst $z^1$.  Then, finally, the worst possible 
behavior of amplitudes as $z\to \infty$ is $z^2$, times the $z$-dependence of
the external gluon polarization vectors.

If we take $q$ as the reference vector for the polarization vectors of the 
shifted gluons, these polarization vectors scale as
\beqa
\epsilon_+^\mu(\hat{i})&=&
{\langle q\gamma^\mu i] \over \sqrt{2}\spa{q}{\hat{i}}}\sim {1 \over z} \ ,
\qquad \epsilon_+^\mu(\hat{j}) 
={\langle q\gamma^\mu\hat{j}] \over \sqrt{2}\spa q j}\sim z\ ,\CR
\epsilon_-^\mu(\hat{i})
&=&-{[q\gamma^\mu \hat{i}\rangle \over \sqrt{2}\spb q i}\sim z\ ,
\qquad\epsilon_-^\mu(\hat{j})
=-{[q\gamma^\mu j\rangle \over \sqrt{2}\spb{q}{\hat{j}}}\sim {1 \over z}\ .
\eeqa{polvectors}
We conclude that, in the three cases of shifts allowed in standard QCD, 
the shifted amplitudes behave at worst as
\begin{center}
\begin{tabular}{c c c}
$\hat{i}$&$\hat{j}$&large $z$\\
\hline
$+$&$+$&$z^2$\\
$+$&$-$&$1$\\
$-$&$-$&$z^2$
\end{tabular}
\end{center}

However, the true situation is slightly better.  For an anomalous 
magnetic moment vertex that stands in front of a fermion propagator 
carrying the shift momentum,
\beq
        ga  \Sigma^{\mu\nu} (zq_\nu + k_\nu)  {z\qslash + \kslash' + m\over
              (z q + k')^2 - m^2} \ ,
\eeq{Sigmaform}
we can rewrite
\beq
   \Sigma^{\mu\nu}q_\nu =  i (\gamma^\mu \qslash - q^\mu) \ .
\eeq{sigmaid}
Since $q^2 = (\qslash)^2 = 0$, the $\qslash$ term cancels the leading 
term in the propagator, and the $q^\mu$ term either vanishes when dotted into
a polarization vector or dots with  a $q$ in a 3-gluon vertex and
thus cancels the leading $z$ term in this vertex.  For a magnetic moment
vertex behind a fermion propagator carrying the shift momentum, a similar
manipulation applies.  This reduces the estimates in the table by at least a factor of
$z^{-1}$.   In this way, we see that the $+-$ shift allows 
a BCFW reduction, while the $++$ and $--$ shifts still may not.

To resolve these last cases, it is simplest to directly compute the 
amplitudes for 2 gluons with one magnetic moment vertex in the case of 
massless fermions.  For massless fermions, \leqn{basicL} is not well-defined.
However, a massless fermion can have an anomalous magnetic moment, and so
we replace $m$ in the denominator of the last term in \leqn{basicL} with some
high scale $M$.  This prescription for massless fermions will also be used 
in the discussion in the Appendix.
In standard QCD, the massless fermion amplitudes with
two $+$ or $-$ helicity gluons vanish.  With nonzero $a$, this is no longer
the case.  We find
\beq
{\cal A}(q^+,g_1^+,g_2^+,\bar{q}^+)= 
           {g^2 a \over 2M} {{\spb 12}^2 \over \spa q {\bar q}} \ .
\eeq{plusplusex}
This expression behaves as $z^0$ after a $++$ shift on the gluons. 
In contrast
\beq
{\cal A}(q^-,g_1^-,g_2^+,\bar{q}^-) = 
- {g^2 a \over 2M}{{\spa q1}^2{\spa {\bar q} 1}^2 \over \spa q1\spa 12\spa 2{\bar q} } \ .
\eeq{minusplusex}
  This behaves as $z^{-1}$ after a $+-$ shift.   This confirms that our 
current estimates are, in general, the best possible.  The BCFW recursion 
can be used to reduce amplitudes for which a $+-$ shift is possible, but,
for $a \neq 0$, 
it cannot be used in the cases of $++$ and $--$ shifts.

Using $+-$ shifts only, we can reduce any amplitude for $q\bar q + n g$
to amplitudes that involve all $+$ helicity gluons or all $-$ helicity
gluons.  However, we cannot, in general, go further.  We need another 
method to compute these cases, which are required as input to the general
 $q\bar q + n g$ amplitude.

For the case of gluons coupling to massless quarks, we have obtained an
explicit formula for the amplitudes with all $+$ helicity gluons.  This is
presented in Appendix A.  We have not succeeded in generalizing this to the 
case of massive fermions relevant for top quark physics.  In the next
section, we will take up another approach to this problem.

\section{An Auxiliary Scalar Theory}\label{phenscalar}

We can make progress toward the computation of the all $+$ helicity gluon 
amplitudes by breaking up \leqn{basicL} into chiral components and 
rearranging it into a second-order Lagrangian.  

Let $(\psi, \bar \psi)$ be the 
left- and right-handed spinor components of $\Psi$, so that
\beq
  \Psi = \pmatrix{ \psi \cr \bar \psi \cr }\ , \qquad
 \bar\Psi = \pmatrix{ \bar\psi^\dagger & \psi^\dagger \cr } \ .
\eeq{Psidecomp}
In this basis, the Dirac matrices take the form
\beq
   \gamma^\mu = \pmatrix{ 0 & \sigma^\mu \cr \bar \sigma^\mu & 0\cr }
\eeq{Diracgamma}
where $\sigma^\mu = (1, \vec \sigma)^\mu$  and $\bar \sigma^\mu = (1 , - \vec
 \sigma)^\mu$, and 
\beq
  \Sigma^{\mu\nu}  =  2i \pmatrix{  \sigma^{\mu\nu} & 0 \cr 
                         0 & \bar \sigma^{\mu\nu} \cr } \ ,
\eeq{Sigmadecomp}
where 
\beq
  \sigma^{\mu\nu} = {1\over 4}(\sigma^\mu \bar \sigma^\nu - \sigma^\nu
           \bar\sigma^\mu) \quad 
  \bar \sigma^{\mu\nu} = {1\over 4}(\bar\sigma^\mu \sigma^\nu 
             - \bar \sigma^\nu \sigma^\mu) \ .
\eeq{littlesig}
The Lagrangian \leqn{basicL} becomes
\beqa
\L &=& \psi^\dagger (i\bar\sigma\cdot D) \psi 
          + \bar\psi^\dagger (i\sigma\cdot D) \bar\psi 
        - m \psi^\dagger \bar\psi - m \bar\psi^\dagger \psi \CR
 & &  \hskip 1.0in   + i {g a\over 2m} \psi^\dagger 
           (\bar \sigma_{\mu\nu}F^{\mu\nu a} t^a)  \bar \psi  
 + i {g a\over 2m} \bar \psi^\dagger 
           (\sigma_{\mu\nu}F^{\mu\nu a} t^a)  \psi\ .
\eeqa{splitL}

Now formally integrate out $\bar\psi$ and $\psi^\dagger$.  This gives
\beq
  \L = \bar\psi^\dagger \biggl[ -m 
               +i {ga\over 2m}\sigma_{\mu\nu}F^{\mu\nu a} t^a
    + \sigma\cdot D {1\over m  - i (ga/2m)\bar \sigma_{\mu\nu}F^{\mu\nu a} t^a}
         \bar \sigma \cdot D \biggr] \psi \ .
\eeq{firstnewL}
After Taylor expanding the denominator and using the properties of the sigma 
matrices, this becomes
\beq
  \L = {1\over m} \bar\psi^\dagger \biggl[ -D^2 - m^2 
               -i {g \hat g_L\over 2}\sigma_{\mu\nu}F^{\mu\nu a} t^a
    - \sigma\cdot D \sum_{n=1}^\infty\biggl( {-iga\over 2 m^2}
\bar \sigma_{\mu\nu}F^{\mu\nu a} t^a\biggr)^n 
         \bar \sigma \cdot D \biggr] \psi \ .
\eeq{secondnewL}
In this equation, the factor $a$ in front of $\sigma\cdot F$ in 
\leqn{firstnewL} has combined with a term arising from the commutator
of covariant derivatives to produce the factor
\beq
       \hat g_L =  2 + a 
\eeq{Landeg}
Thus, we obtain a second-order equation with a term close to the
full magnetic moment of the fermion appearing explicitly.
Note that \leqn{Landeg}
 differs from  the Land\'e $g$ factor of the fermion, which 
would be $(2 + 2a)$.  The Land\'e $g$ factor refers to the behavior of the
fermion in a background magnetic field.  The $\vec\sigma \cdot \vec B$ 
term gets contributions both from $\sigma \cdot F$ and $\bar\sigma \cdot F$.
Thus, the $n=1$ term in the sum in the last term also contributes to the 
Land\'e $g$ factor, supplying the missing contribution of $a$.

If we had chosen instead to integrate out $\psi$ and $\bar\psi^\dagger$, we 
would have obtained the same second-order action with the positions of 
$\sigma^{\mu\nu}$ and $\bar\sigma^{\mu\nu}$ interchanged.  The significance
of this exchange will be made clear below.

The procedure of integrating out components of the quark 
field is used in other contexts in infinite momentum frame 
quantization~\cite{infmom}, light cone QCD~\cite{lightcone}, and 
soft and collinear effective field theory~\cite{SCET}. 
For the application here, we would like to emphasize that this integration
out introduces no approximations.  From \leqn{secondnewL}, we are able
to reconstruct any amplitude in the original theory.   Although our 
new Lagrangian is not the most convenient way to obtain the scattering
amplitudes in the limit $m\to 0$, it does give the correct answers
in this limit, as
we will illustrate in Section 5.

To analyze the consequences of \leqn{secondnewL}, it is tempting to drop
the series of terms with $\bar\sigma\cdot F$ and approximate this theory 
by 
\beq
  \L = {1\over m} \bar\psi^\dagger \biggl[ -D^2 - m^2 
               + i {g \hat g_L\over 2}\sigma_{\mu\nu}F^{\mu\nu a} t^a
                                        \biggr] \psi \ .
\eeq{newL}
This theory resembles a relativistic theory of a scalar field, except that this
scalar retains a 2-component internal spin variable on which 
$\sigma^{\mu\nu}$ acts. In the following, we will refer to this model as a 
scalar theory even though it does describe spin $\half$. 
 To better understand the relation of this theory
to the original Dirac theory, note that if we start from the Dirac equation
with $a = 0$
\beq
  (i\Dslash - m) \Psi = 0 
\eeq{originalDirac}
and multiply by $(i\Dslash + m)$ on the left, we obtain 
\beq
   (-D^2 - m^2 + {g\over 2} \Sigma_{\mu\nu} F^{\mu\nu a} t^a)\Psi = 0 \ ,
\eeq{secondorderDirac}
in which the top two components are precisely the equation of motion 
from \leqn{newL} with $\hat g_L = 2$.  The equation \leqn{secondorderDirac}
arises in calculating of the determinant of the Dirac operator, for example,
in the background-field derivation of the QCD beta function.

In general, there is no justification for approximating \leqn{secondnewL} 
by \leqn{newL}.  However, we are interested here in computing the 
amplitude for $q\bar q$ plus gluons with $+$ helicity only.  A configuration
of gluons with all $+$ helicities is a self-dual Yang-Mills 
field~\cite{selfdual}.  The operator $\sigma_{\mu\nu}F^{\mu\nu}$ projects
onto self-dual field configurations.  Conversely, 
 $\bar\sigma_{\mu\nu}F^{\mu\nu}$ projects onto anti-self-dual configurations
and is zero in a self-dual background~\cite{explicit}.  So, precisely
for the situation of computing an amplitude with all $+$ helicity gluons, we
may use \leqn{newL} as a replacement for \leqn{secondnewL}, with is 
equivalent to \leqn{basicL}.   The same argument implies that, for 
computing amplitudes with all $-$ helicity gluons, we may use the second-order
Lagrangian obtained by integrating out $\bar\psi^\dagger$ and $\psi$, 
which has the form of \leqn{newL} with $\sigma^{\mu\nu}$ replaced by 
$\bar\sigma^{\mu\nu}$.

 The Feynman rules for the theory \leqn{newL} are the same as those for
scalar QCD,  augmented with the new vertices from the magnetic moment 
term. These vertices contain $2\times 2$ sigma matrices which must be 
evaluated in the correct external spin states.  To compute Feynman diagrams
in this theory, we first evaluate the diagrams as in a scalar theory with 
an internal spin.  The sum of  diagrams 
will contain a product of $\sigma^{\mu\nu}$
matrices.  We must then take the matrix element of this product using
the 2-component spinor corresponding to the components of $\bar u$
or $v$ in \leqn{qspinors}, \leqn{qbarspinors} 
that have not been integrated out.  Specifically, to compute 
the amplitude for an outgoing fermion with momentum $p$, in a spin 
basis described by the reference vector $r$, we use the spinors
\beq
  \mbox{for}\ q_R\ : \ [p^\flat  \ , \qquad  \mbox{for}\ q_L\ : \ { [r\, m
         \over [rp^\flat]} \ ,
\eeq{twoqspinors}
where $p^\flat$ is defined by \leqn{pflatdef}.  Similarly, for an outgoing
antifermion, we use
\beq
  \mbox{for}\ \bar q_R\ : \ p^\flat] \ , \qquad 
          \mbox{for}\ \bar q_L\ : \ { -m \, r] \over [p^\flat r]}  \ .
\eeq{twoqbarspinors}
A separate reference vector can be used for each external
momentum. Finally, to account the factor $(1/m)$ in front of \leqn{newL}, the 
entire amplitude should be multiplied by $(1/m)$.

At any point, we can break up the products of $\sigma^{\mu\nu}$ matrices
using the following completeness relation:  Let $a$, $b$ be any lightlike
vectors that are not collinear.  Then
\beq
             1 =   {   a][ b - b] [ a \over [ba]}\ .
\eeq{twocompleteness}
The object on the right is the identity when acting on $a]$ and $b]$, which
are independent 2-component vectors, so it must be the identity in general.
This formula is useful to write the right-hand side of the BCFW identity
as a pair of amplitudes in the scalar theory.  On a cut line, the scalar
4-momentum will be on-shell, but \leqn{twoqspinors}, \leqn{twoqbarspinors}
will be replaced by the more arbitrary spinors $a]$ and $b]$.

For this construction to be truly useful, we need to show that it allows
a BCFW recursion that computes the scalar amplitudes in the case of all $+$ 
helicity gluons.  For this scalar theory, we can show that the $+-$ 
gluon shift is always allowed by na\"{i}ve power counting alone. 
 First, if the shifted gluons are only separated by gluon propagators,
 the good large $z$ behavior is guaranteed by arguments from pure QCD.
  The only issues arise if the deformed momentum flows through a scalar
 propagator.  To argue that even in these cases, the large $z$ behavior 
is good, we note that scalar propagators scale as $1/z$ and that trees of 
gluons which contain one of the shifted gluons also scale as  $1/z$. 
 The usual scalar QCD vertices scale at worst as $z$ for 
 large $z$.  Thus, the $+-$ gluon shift is always allowed in scalar
 QCD.  The new, helicity violating vertices also scale at worst as $z$
 for large $z$, since they are also proportional to momentum, and so the
 $+-$ gluon shift is always allowed in this theory.  However, this
 argument fails for the $++$ gluon shift and so the scalar theory
has the same problem 
 that we found in the fermion theory.

\begin{figure}
\begin{center}
\includegraphics[height=1.5in]{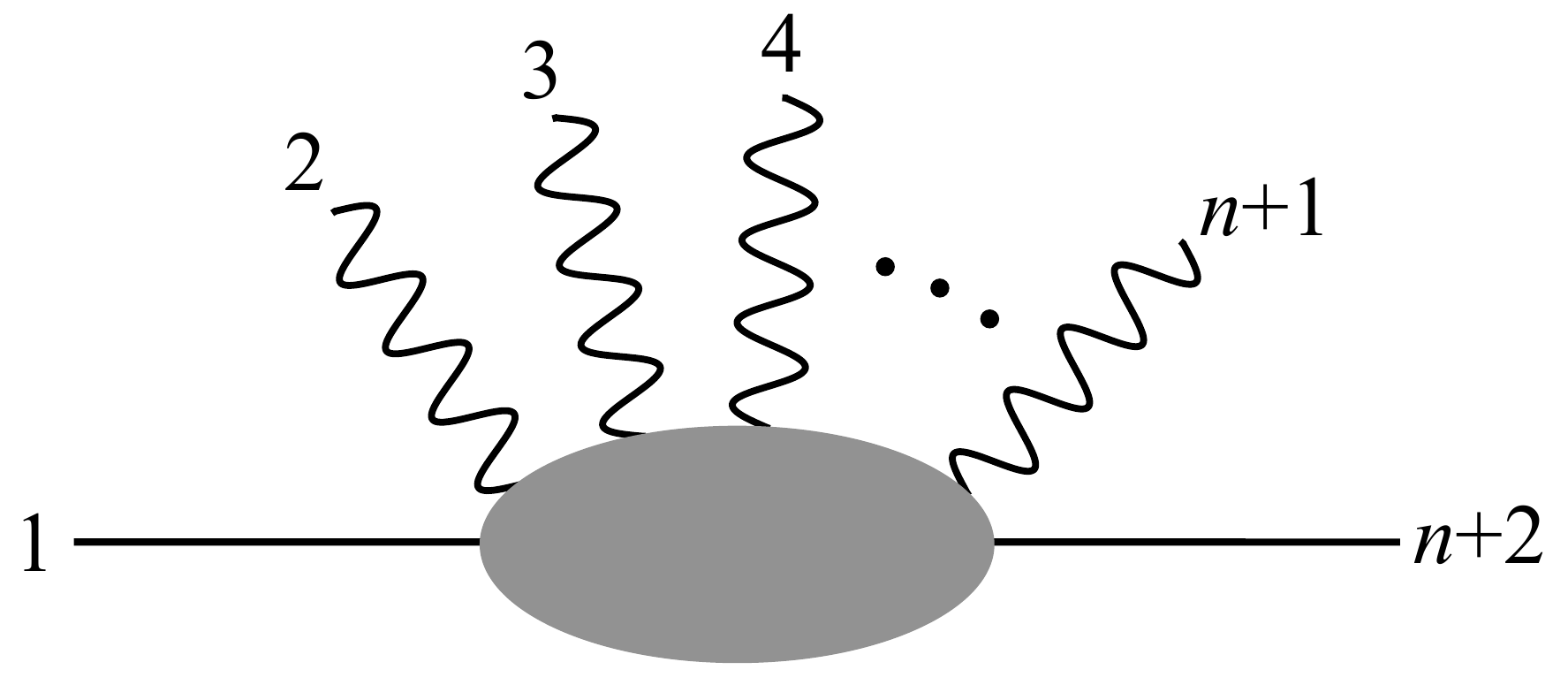}
\caption{Notation for the $q\bar q + n g $ color-ordered amplitude.}
\label{fig:basicscalar}
\end{center}
\end{figure}

However, there is another possible shift in the scalar theory 
that is allowed in the case of all $+$ gluon amplitudes.
This involves a gluon and an external scalar.  Consider the 
$q\bar q + n g$ amplitude with the external legs numbered as in
Fig.~\ref{fig:basicscalar}.  When the  gluon $3$ is shifted by 
\beq 
3\rangle\to 3\rangle-z\; l_1\rangle \ ,
\eeq{gluonshift} 
where $l_1$ is lightlike,
 and the external scalar $1$ is shifted by 
\beq
 1\to 1+z \; 3] \langle l_1 \ , 
\eeq{scalarshift}
 we will show that the all $+$ amplitude scales as 
 $1/z$ for large $z$.  Particle 1 is massive, and $l_1$ must be defined
so that the shifted vector 1 remains on mass shell.  This requires
\beq
\hat{1}^2= (1+z\; 3]\langle l_i)^2=  1^2 + z [3\, 1\, l_1\rangle = 1^2 \ ;
\eeq{squarehat}
that is,
\beq
               [3 \, 1\, l_1\rangle = 0 \ .
\eeq{conditiononellone}
A general solution to this condition is
\beq
            l_1 =   1 -  {m^2 \over [3\, 1\, 3\rangle} 3 \ ,
\eeq{solutionellone}
that is, that $l_1$ is the flatted vector of 1 computed using 3 as a 
reference vector.

In the fermion theory, this shift could also be defined, but it would restrict
the choice of the reference vector for the external fermion 
to $r = 3$~\cite{Schwinn}.  In the scalar theory, we are free to make this
choice of a shift without any restriction on the reference vector that
will appear in \leqn{twoqspinors}.

\begin{figure}
\begin{center}
\includegraphics[height=1.8in]{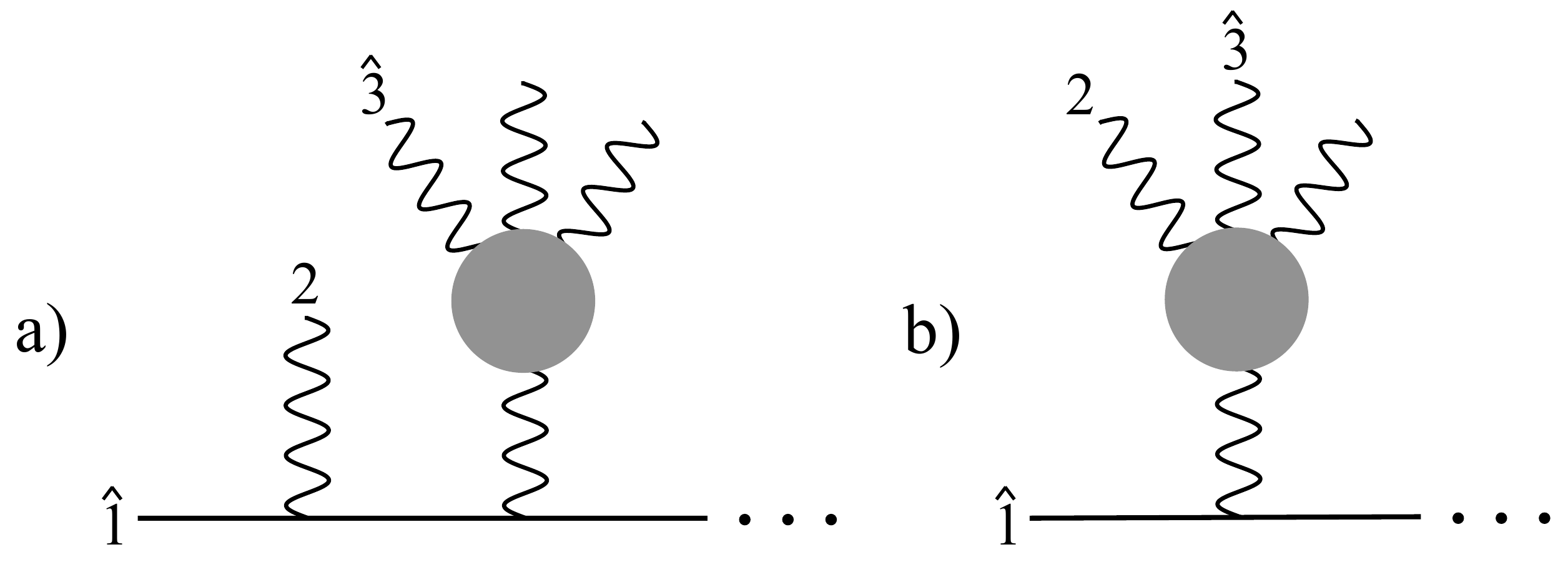}
\caption{Two classes of diagrams needed to analyze the large $z$ behavior
of the 13 shift of the amplitude shown in Fig.~\ref{fig:basicscalar}.}
\label{fig:twocases}
\end{center}
\end{figure}

We will now prove that the above shift on the legs 1 and 3 has the good
$z\to \infty$ behavior that we have claimed. To do this, we 
consider the possible forms of Feynman diagram that can contribute
 to an amplitude in this theory with arbitrary numbers of helicity
 violating vertices.  We need only consider the left-most piece of
 the diagram that contains the scalar line 1 and the gluons 2 and 3.
For this, there are 
two possibilities: either gluon 2 is connected directly to the scalar 
line and gluon 3 is part of a tree of gluons with external legs $ 3 \cdots k$,
or gluons 2 and 3 are part of the same tree of gluons with external legs 
$2 \cdots k$. These cases are shown in Figs.~\ref{fig:twocases} (a) and (b).

To analyze the trees, we need the explicit expresson for the
tree in the case in which all gluons have $+$ helicity.  This expression,
for the case in which all gluons have the same reference vector $r$, is
worked out in the Appendix. The result, for external gluons $j \cdots k$,
is
\beq
     J_a(j,\cdots, k) =  -i {(j + \cdots + k) r\rangle \langle r (j + \cdots
          + k)\over \spa rj \spa j {(j+1)} \cdots \spa {(k-1)} k \spa kr } \ .
\eeq{Jexpression}
The dangerous term here will be the one that involves the shifted momentum
$\hat 3$ in both terms of the numerator.  Note that $\hat 3 $ also appears
twice in the denominator, so this term is only of order $z^0$.

Now consider the first case, shown in Fig.~\ref{fig:twocases}(a).  
The scalar propagator scales as $z^{-1}$.  If we take the gluon 2 to couple
via a magnetic moment vertex, the vertex will be of order 1 and the diagram
will vanish as $z\to \infty$. Thus, the only dangerous diagram is that in
which 2 couples by the ordinary scalar vertex.  The leading term in this 
diagram is
\beq
   i {\langle r \hat 1 2]\over \spa r2} { i\over \langle 2 \hat 1 2]}
       {-i\ \hat 3 r\rangle \langle r \hat 3 \over \spa r{\hat 3} \spa {\hat 3} 4
          \cdots \spa kr} \cdots \ .
\eeq{casea}
Notice that the numerator of the last term is a matrix.  This is a 
$2\times 2$ sigma matrix for which we must eventually take the matrix
element between the spinors \leqn{twoqspinors} and \leqn{twoqbarspinors}.
This term scales as $z^0$.

In the second case, the value of the first part of the diagram is just
that of the tree \leqn{Jexpression}.  The dangerous term is
\beq
      -i {\hat 3 r\rangle \langle r \hat 3 \over \spa r2 
               \spa 2 {\hat 3} \spa {\hat 3} 4
          \cdots \spa kr} \cdots  \ .
\eeq{caseb} 

These two bad pieces contribute to the amplitude at the same order
of $g$ and $a$, and so we may add them together.  Then an 
amazing thing happens. The sum is 
\beq
    -i {\hat 3 r\rangle \langle r \hat 3 \over \spa r2 \apb 2 {\hat 1} 2
        \spa 2{\hat 3} \spa r{\hat 3} \spa {\hat 3} 4
          \cdots \spa kr} \cdot [ \langle 2 \hat 1 2] \spa r{\hat 3}
       -  \langle r \hat 1 2] \spa 2 {\hat 3} ]  \cdots \ .
\eeq{sumofcases}
The quantity in brackets is
\beq
    \bpa 2{\hat 1}2 \spa r{\hat 3} - \bpa 2{\hat 1} r
       \spa 2 {\hat 3}  = 
 \spa r2   \bpa 2{\hat 1}{\hat 3} \ , 
\eeq{moresum}
by the Schouten identity.  Since the leading $z$ term in $\hat 1$ and $\hat 3$
is proportional to the same lightlike vector, this term cancels in the 
last product.  Then the sum of diagrams scales as $z^{-1}$ and the 
sum has good behavior as $z\to \infty$.
  This proves our claim that the 
shift on 1 and 3 generates a BCFW recursion formula for the amplitude 
with all $+$ helicity gluons.

We now have an algorithm for computing any $q\bar q + ng$ amplitude for 
nonzero $a$.  If the amplitude contains both $+$ and $-$ helicity gluons, 
we can apply $+-$ shifts of the gluons to reduce the amplitude to lower-point
components.  If the amplitude has only $+$ helicity gluons, we can use the
13 shift above in the scalar theory to reduce the amplitude to lower-point
components.  If the amplitude has only $-$ helicity gluons, we can use the
scalar theory with $\bar \sigma\cdot F$.  In this theory, a 13 shift that
shifts the square bracket of 3 reduces the amplitude to lower-point 
components.  Eventually, the recursion gives the original amplitude in terms
of on-shell three-point amplitudes.  Though we have given the argument
explicitly only for \leqn{basicL}, the same strategy works when the 
QED anomalous magnetic moment interaction \leqn{QEDanom} is added to the 
theory.

\section{Calculations in the Scalar Theory}\label{calcscalar}

Although we have shown that the scalar theory described by \leqn{secondnewL}
or \leqn{newL} can be effective for computing $q\bar q + ng$ amplitudes,
some aspects of this theory still appear odd.  Of these, the oddest feature
is the factor of $1/m$ in front of the Lagrangian.  Some diagrams will then
contain factors of $1/m$, and one might worry that these would generate 
bad behavior in the limit $m\to 0$.  In this section, we will display some
amplitudes in the theory \leqn{newL} that might provide sanity checks
on the use of that expression.

First, consider three-point amplitudes.  The scalar amplitude as a 
$2\times 2$ matrix is 
\beq
{\cal A}_s(1,2^+,3)= {g \over m} \biggl[ 
{\apb r12\apb s32 \over \langle s31r\rangle+m^2\spa sr}+{\hat g_L \over 2 }2][2 
    \biggr] \ ,
\eeq{threepointscalar}
where $r$ and $s$ are the reference spinors for particles 1 and 3, 
respectively.  Setting $\hat g_L=2$, the fermion amplitudes can be computed 
by taking matrix elements in \leqn{twoqspinors} and \leqn{twoqbarspinors}.
We find
\beqa
{\cal A}(1^+,2^+,3^+)&=&{gm \over \spa{r}{1^\flat}\spa{3^\flat}{s}}
{\spa sr\apb r12\apb s32 \over \langle s31r\rangle+m^2\spa sr} \ ,\\
{\cal A}(1^+,2^+,3^-)&=&-g{\spb{1^\flat}{2}^2 \over
\spb{1^\flat}{3^\flat}}-{g m^2 \over \spa{r}{1^\flat}\apb{s}{3}{s}}
{\spa rs \spb{1^\flat}{s} \over \spb{1^\flat}{3^\flat}}
{\apb r12\apb s32 \over \langle s31r\rangle+m^2\spa sr} \ ,\\
{\cal A}(1^-,2^+,3^+)&=&g{\spb{2}{3^\flat}^2 \over
\spb{1^\flat}{3^\flat}}+{gm^2 \over \apb{r}{1}{r}\spa{3^\flat}{s}}
{\spa rs\spb{r}{3^\flat} \over \spb{1^\flat}{3^\flat}}
{\apb r12\apb s32 \over \langle s31r\rangle+m^2\spa sr} \ ,\\
{\cal A}(1^-,2^+,3^-)&=&-gm{\spb r2\spb 2s \over \spb{r}{1^\flat}
\spb{3^\flat}{s}}-gm{\spb rs \over \spb{r}{1^\flat}\spb{3^\flat}{s}}
{\apb r12\apb s32 \over \langle s31r\rangle+m^2\spa sr} \ .
\eeqa{randomeq}
These expressions are in agreement with  explicit QCD calculations.  
Taking the limit $m\to 0$,
 these expressions reduce to the familiar three point maximal 
helicity violating (MHV) amplitudes
\beqa
{\cal A}(1^+,2^+,3^-)&=&-g{\spb{1}{2}^2 \over\spb{1}{3}} \ ,\\
{\cal A}(1^-,2^+,3^+)&=& g {\spb{2}{3}^2 \over \spb{1}{3}}  \ ,
\eeqa{randomtwo}
with ${\cal A}(1^+,2^+,3^+)= {\cal A}(1^-,2^-,3^-)= 0$.

At four points, there exist two helicity configurations of the 
gluons that cannot be related by parity.  
These amplitudes can be computed in the scalar theory; we find
\beqa
{\cal A}_s(1,2^+,3^+,4) &=& g^2m {[32] \over
\langle 2123\rangle}+{g^2 \hat g_L \over 2m}\biggl({m^2 \ 2][3 \over
\langle 2123\rangle}+{12\rangle\spb 32[2 \over 
\langle 2123\rangle}+{412][3 \over \langle 2123\rangle}
+{\hat g_L \over 2}{ 2][23] [3 \over \apb 212}\biggr)\CR
{\cal A}_s(1,2^+,3^-,4) &=&-{g^2 \over m}{\apb 312 ^2 \over
\apb 212\apb 232}-{g^2 \hat g_L \over 2m}
{\apb 312 \over \apb 212\spb 23} 2][2  \ .
\eeqa{scalaramp}
To compare to fermion amplitudes, we need to take matrix elements of these
$2\times 2$ matrices.
For brevity, we will only consider the $\hat g_L=2,$ $m=0$ case.  
For the case with both gluons with $+$ helicity, the massless 
fermion amplitude with any helicity configuration 
for the fermions must vanish.  For massless fermions, 
the  $--$ fermion projection explicitly vanishes. 
 Multiplying this scalar amplitude by $[a$ on the left 
and $b]$ on the right and simplifying yields
\beq
[a\ {\cal A}_s(1,2^+,3^+,4)\ b]
= {g^2 \over m}{\spa 41\spb 23 \spb 4b\spb 1a \over
\apb 212\spa 23} \ .
\eeq{4ptmasslesswrong}
This indeed vanishes if either $1$ or $4$ have $+$ helicity.

In the second case, in which 
 the gluons have opposite helicity, the projection should yield the 
familiar MHV amplitudes at four points.  If we choose both particles 
1 and 4 to have $+$ helicity, the projection vanishes by momentum 
conservation.  If instead, particles 1 and 4 have opposite helicity, 
the projection yields
\beqa
{\cal A}(1^+,2^+,3^-,4^-)&=&g^2 {\spa 13\spa 34 ^3 \over
\spa 12\spa 23\spa 34\spa 41} \ ,\\
{\cal A}(1^-,2^+,3^-,4^+)&=&g^2 {\spa 13 ^3\spa 34 \over
\spa 12\spa 23\spa 34\spa 41} \ ,
\eeqa{twopointex}
which agree with the standard results.

As discussed in the previous section, 
the all $+$ helicity amplitudes are completely described by this theory.  
In fact, from the amplitude in \leqn{scalaramp}, one can verify that, 
order by order in $a$, this expression agrees with
 that calculated using from \leqn{basicL}.  However, 
a simple observation on the opposite helicity amplitude in \leqn{scalaramp}
 shows that this 
amplitude cannot reproduce the full result from \leqn{basicL}.  The result 
above contains only terms proportional to $a^0$ and $a^1$, 
while the exact answer would also contain a term proportional to $a^2$.
  This discrepancy is expected, and it is not troublesome for us, since
 this amplitude in the original theory can be constructed using BCFW directly.

There is one more interesting cross check that we have made of the form of
\leqn{newL}.  For $\hat g_L = 2$, $a =0$, and so \leqn{newL} gives an exact
description of \leqn{basicL} for all gluon helicity states.  At the same time,
for $a =0$, all amplitudes of \leqn{basicL} can be computed by BCFW shifts
on gluons with the helicity combinations $+-$, $++$, $--$.  Thus, one can 
compute every $q\bar q + ng$ amplitude in two ways, first, from \leqn{basicL}
using gluon shifts only and, second, from \leqn{newL}, using the 13 shift
described in the previous section.  We have checked equality numerically 
to 6 significant figures for all of these amplitudes up to $n=8$ gluons.

\section{Conclusion}

We have shown that the BCFW recursion relations can be used to 
compute all amplitudes in a theory with an anomalous magnetic moment.  
The prescription for using BCFW is as follows:
\begin{enumerate}
\item If an amplitude contains at least one $-$ and one $+$ helicity gluon, 
use the $+-$ shift to compute amplitudes in the theory defined by
\beq
{\cal L}=\bar{\Psi}(i\Dslash -m)\Psi
+{ga \over 2m}\bar{\Psi}\Sigma_{\mu\nu}F^{\mu\nu a}t^a\Psi \ .
\eeq{finalone}
\item If an amplitude contains only $+$ 
helicity gluons, shift on the scalar and a non-adjacent gluon 
to compute amplitudes in the theory defined by
\beq
{\cal L}={1\over m}\bar{\psi}^\dagger\biggl[-D^2-m^2+i{g \hat g_L \over 2}
\sigma_{\mu\nu}F^{\mu\nu a}t^a\biggr]\psi \ .
\eeq{finaltwo}
The gluon momentum should be shifted in the angle bracket.
To compute the amplitude with external fermions, project onto the fermion 
line by multiplying by the appropriate wavefunctions on the left and right.
\item If an amplitude contains only $-$ 
helicity gluons, shift on the scalar and a non-adjacent gluon 
to compute amplitudes in the theory defined by
\beq
{\cal L}={1 \over m}\bar{\psi}\biggl[-D^2-m^2+i{g \hat g_L \over 2}
{\bar \sigma}_{\mu\nu}F^{\mu\nu a}t^a\biggr]\psi^\dagger \ .
\eeq{finalthree}
The gluon momentum should be shifted in the square bracket.
To compute the amplitude with external fermions, project onto the fermion 
line by multiplying by the appropriate wavefunctions on the left and right.
\end{enumerate}
We have shown that this is an efficient algorithm for computation of 
tree amplitudes.  We hope to present some phenomenological applications of
this method soon.

Our conclusions include the statement that the BCFW recursion formula
cannot be used to fully construct amplitudes in
 the original fermion theory.  This apparently contradicts a result 
of~\cite{cliff}, although in fact the anomalous magnetic moment coupling
falls outside the hypotheses of that paper~\cite{cliffemail}.
More generally, the validity of BCFW recursion must be thought through
carefully for effective theories with nonrenormalizable couplings.
However, our analysis indicates that remedies for their bad large-momentum
behavior can be found in some cases.

A distinct momentum shift useful for studying generic theories was introduced 
in \cite{CEK}.  Instead of only shifting 
the momenta of two of the particles in an amplitude, the authors consider 
shifting the momentum of all external particles.  Explicitly, for an amplitude 
with all massless particles, the shift can be expressed as
\beq
i\rangle\to  i \rangle + w_i \ z \ X\rangle.
\eeq{CEKshift}
$i$ is any external particle in the amplitude, $X$ is an arbitrary, massless 
four-vector and the coefficients $w_i$ are chosen to conserve momentum:
\beq
\sum_i w_i \ i] = 0.
\eeq{wsum}
The dependence on the parameter $z$ is easily 
determined by considering the dimension and helicity 
constraints on an amplitude in a generic theory.  For the case 
of the shift in \leqn{CEKshift}, the amplitude behaves as
\beq
{\cal A}\sim z^s \mbox{ as } z\to \infty \mbox{ with } 2s = 4-n-c-H,
\eeq{zdep}
where $n$ is the number of external legs, $c$ is the sum of 
dimensions of coupling constants
in an amplitude and $H$ is the sum of helicities of external 
particles.  An on-shell recursion exists 
when $s<0$; in that case, there are more angle brackets in the denominator 
of an amplitude than in the 
numerator.  In the anomalous magnetic moment theory, this 
all-leg shift leads to a 
recursion relation precisely for those amplitudes for which 
BCFW fails.  This is easily seen 
at the four point level from \leqn{plusplusex} 
and \leqn{minusplusex}.  In \leqn{plusplusex}, 
the amplitude is constructible with this shift because 
there is one angle bracket in 
the denominator and none in the numerator, while 
in \leqn{minusplusex} there is one more
 angle bracket in the numerator and so this amplitude 
is not constructible.  This all-leg shift 
could be another way to compute amplitudes in a theory 
with an anomalous magnetic moment.  
Unfortunately, its practical use is limited because of 
the proliferation of cuts that one 
needs to compute.

Recently, there has been some interest in the literature in finding
classes of theories in whose amplitudes are 
constructible using BCFW \cite{freddy,elvang}.  The  hope has been that
the validity of the BCFW recursion formula would
 say something about the behavior
of the  theory at high energy. It is remarkable that amplitudes in 
Einstein gravity and, equivalently, in ${\cal N}=8$ supergravity, 
are  are constructible 
using BCFW \cite{jared,gravbcfw}.   It has been hoped that this property
is evidence for special simplicity of the ${\cal N}=8$ theory.  Further speculations
on this point  ought to take into account, one way or the other, our result
that QCD with an anomalous magnetic moment is also BCFW constructible.
  We hope that the methods 
discussed here can be used to study other realistic or effective theories,
and that those investigations will shed more light on the high-momentum
behavior of non-renormalizable theories.

\appendix

\section{All-$+$ gluon helicity amplitude for massless quarks}

For massless particles, we have found an explicit formula for the
$q\bar q + ng$ amplitudes with all $+$ helicity gluons. 
The derivation of this formula makes use of the  off-shell 
current formalism of Berends and Giele~\cite{BG}.  The off-shell
current with all $+$ helicities is needed for other arguments
in this paper, in particular, in the analysis of the large $z$ 
behavior of the scalar theory at the end of Section 4.

We consider
an amplitude with a single massless fermion line and $n$ $+$ helicity 
gluons.  We would like to compute the correction to the standard QCD
result coming from the presence of an anomalous magnetic moment.
  Since the background field with only $+$ helicity gluons is
self-dual, the $\bar \sigma \cdot F$ piece of the magnetic moment
operator gives zero and only the $\sigma\cdot F$ of this operator
contributes to the amplitude.  This term has a matrix element only between
a $+$ helicity fermion and a $+$ helicity antifermion.   All
propagators and all other vertices in the diagram are helicity-conserving.
This means that, for the
 amplitude to be non-zero, the helicity of both external 
fermions must be $+$ and there must be exactly  one insertion of the 
magnetic moment operator.

The magnetic moment operator 
contains both a three-point and a four-point
vertex.  Both terms contribute to the Berends-Giele current.  The 
terms simplify, however, if we choose all of the $+$ helicity gluons to 
have the same reference vector $r$.  Our analysis here generalizes the 
results of Berends and Giele~\cite{BG} obtained for currents with 
standard QCD vertices.

Consider first the term from the four-point vertex.  Computing the first
few trees emanating from the four-point vertex, we find
\beqa
J_4(1,2)&=&i{(1+2)^2 \over \spa r1\spa 12\spa 2r}\:r\rangle\langle r\ ,\\
J_4(1,2,3)&=&i{(1+2+3)^2 \over \spa r1\spa12\spa23\spa3r}\:r\rangle\langle r\ .
\eeqa{Jfour}
From these expressions, we postulate the general form of this term:
\beq
J_4(1,\ldots,n)=i{(1+\cdots+n)^2 \over \spa r1\spa12\cdots\spa nr}
\:r\rangle\langle r \ .
\eeq{fourptcurrent}
To prove this, note that the BCFW recursion is valid for shifting on any 
two gluons.  Thus, to prove the general expression, shift gluons 2 and 3 
and use induction.  Only one term contributes in the BCFW sum and it is 
given by the above equation.  

Similarly, one can compute the first few all $+$ gluon trees that emanate 
from the three point helicity violating vertex:
\beqa
J_3(1)&=&-i 1][1 = -i{(1)r\rangle\langle r(1) \over
\spa r1 \spa 1r}\ ,\\
J_3(1,2)&=&-i{(1+2)r\rangle\langle r(1+2) \ +\ (1+2)^2\; r\rangle\langle r \over
\spa r1 \spa 12\spa 2r}\ ,\\
J_3(1,2,3)&=&-i{(1+2+3)r\rangle\langle r(1+2+3)\ +\ (1+2+3)^2\; 
r\rangle\langle r \over \spa r1\spa 12\spa 23\spa 3r}\ .
\eeqa{lowthree}
These suggest a general form,
\beq
J_3(1,\ldots, n)=-i{(1+\cdots +n)r\rangle\langle r(1+\cdots +n) + 
(1+\cdots+n)^2\; r\rangle \langle r \over \spa r1\spa 12\cdots\spa nr} \ .
\eeq{threeptcurrent}
and that can again be established by induction.

Note that the first term in the numerator in \leqn{threeptcurrent} gives a 
matrix element between a $+$ helicity fermion and  $+$ helicity antifermion.
The second term gives a matrix element between a $-$ helicity fermion and 
a $-$ helicity antifermion.  However, this latter term, proportional to the
total mass of the gluons in the tree, cancels neatly against 
\leqn{fourptcurrent}.  Finally, we find
\beq
J_a(1, \ldots, n) = 
J_3(1,\ldots, n)+J_4(1,\ldots, n)=-i{(1+\cdots +n)r\rangle\langle 
r(1+\cdots +n) \over \spa r1\spa 12\cdots\spa nr} \ .
\eeq{effhv}

This result for the Berends-Giele current of all $+$ helicity gluons is
quoted in \leqn{Jexpression} and forms the basis for our analysis at the 
end of Section 4.

\begin{figure}
\begin{center}
\includegraphics[height=1.85in]{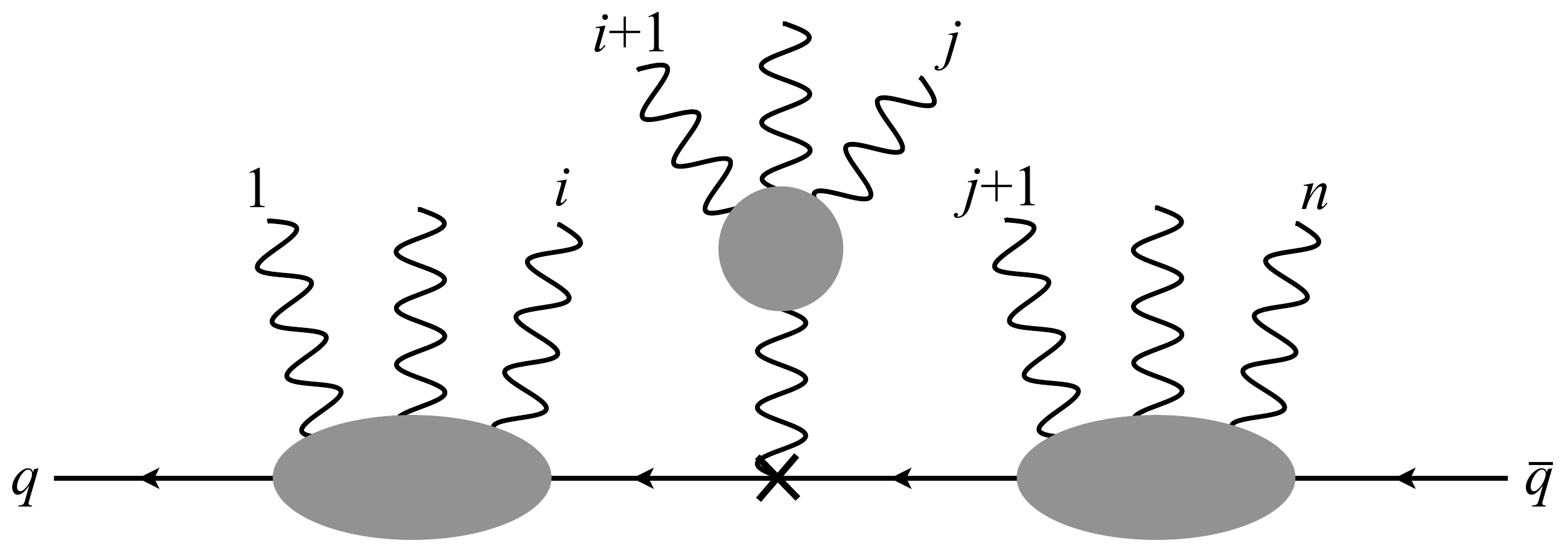}
\caption{Generic form of a $q\bar q + ng$  amplitude for massless
 fermions with a magnetic moment vertex and all $+$ helicity gluons.
 The cross represents the magnetic moment vertex.}
\label{fig:bgfactor}
\end{center}
\end{figure}

In the case of massless QCD with a single helicity violating vertex, 
the amplitude has the diagrammatic form shown in Fig.~\ref{fig:bgfactor}.
We can write the gauge invariant amplitude as
\beq
{\cal A}(q^+,g_1^+,\ldots,g_n^+,\bar{q}^+)={g^n a \over 2M}
\sum_{0\leq i<j\leq n}\Jslash(q;1,\ldots,i)\Jslash_a(i+1,\ldots,j)
\Jslash(j+1,\ldots,n;\bar{q}) \ .
\eeq{masslessallplus}
Here, the factors $J$ are currents with an off-shell fermion leg and 
any number of gluons.  These were first found by Berends and Giele 
in~\cite{BG}.   The factor $J_a$ is the current from a magnetic moment
vertex discussed above.  The explicit forms for the currents $J$ are
\beq
\Jslash(q;1,\ldots,i)=-{\langle r(1+\cdots i+q) \over \spa q 1
\spa 1 2\cdots\spa i r}
\eeq{mlferBG}
and 
\beq
\Jslash(j+1,\ldots,n;\bar{q})=-{((j+1)+\cdots+n+\bar{q})r\rangle \over
\spa{r}{(j+1)}\cdots\spa{n}{\bar{q}}} \ . 
\eeq{mlferrightBG}
The amplitude we wish to compute is
 the sum over all possible insertions of the helicity violating vertex
 with $j-i$ gluons off of the helicity violating vertex and $i$ and 
$n-j$ to the left and right of the helicity violating vertex, respectively.

Plugging in the various pieces, the amplitude becomes
\beq
{\cal A}(q^+,g_1^+,\ldots,g_n^+,\bar{q}^+)= 
{g^n a \over 2M}\sum_{0\leq i<j\leq n}\!\!
{\langle r(q+1+\cdots+i)((i+1)+\cdots+j)r\rangle^2 \over
\spa q 1 \cdots\spa ir \spa{r}{(i+1)}\cdots 
\spa j r\spa{r}{(j+1)}\cdots \spa{n}{\bar{q}}} \ ,
\eeq{qcdzero}
where momentum conservation has been used.  For $n=2$, 
this equation agrees with the expression in \leqn{plusplusex}. 
 This expression is gauge invariant as well.  It is important to note
that gauge invariance follows only after summing over all possible 
places of 
 insertion of the magnetic moment vertex.  In massless QCD,
 this is the end of the story.  We have explicitly constructed the 
amplitude with all $+$ helicity gluons and on all other 
amplitudes, one can use BCFW to construct amplitudes.

Since the amplitude in \leqn{qcdzero} involves only $+$ helicity gluons,
it is also possible to look at this amplitude as a solution for the 
motion of a massless fermion in a purely 
self-dual background field.  For such backgrounds, Rosly and 
Selivanov~\cite{perturbiner} have developed a special formalism, called
the perturbiner method, for amplitude computations.  In this method, 
they solve the Yang-Mills equations recursively in the number of gluons,
then use that solution to evaluate the fermion propagator.  Their
solution can be written as 
\beq
A_\mu=i\sum_{i=0}^\infty{\langle r\gamma_\mu (1+2+\cdots+i)r\rangle \over
\spa r1\spa 12\cdots\spa ir}{\cal E}_1{\cal E}_2\cdots {\cal E}_i \ ,
\eeq{pertA}
for color-ordered gluons, with  $r$  the common 
reference vector for all of the gluons. 
 The objects ${\cal E}_n$ are the solutions to the free equations of motion, 
$${\cal E}_n=a_n t_n e^{ik_n\cdot x} \ ,$$
where $a_n$ is the nilpotent creation operator and $t_n$ is the color matrix. 
 The coefficients of the product of ${\cal E}$s are the Berends-Giele 
off-shell currents for all $+$ helicity gluon configurations.  Applying
\leqn{pertA} to the magnetic moment vertex gives an alternative derivation
of \leqn{effhv}.

This analysis becomes much more complex in the case of massive fermions.
In the massive case, the fermion propagators now have helicity-violating
factors, and so we can insert 
any number of magnetic moment vertices into an amplitude.  
In principle, we can still construct the all $+$ gluon amplitude 
with the stitching procedure used in the massless case.  However, 
to do this, we need to know an explicit form for the analog of 
the off-shell current in \leqn{mlferBG} for massive fermions. 
 In addition, we would need to know this current for both helicities 
of the massive fermion.  We do not show them here as their form is 
not illuminating.
%For small numbers of gluons, this current 
%can be calculated and one finds
%\beqa
%\Jslash(Q^+)&=&[Q^\flat,\\
%\Jslash(Q^+;1)&=&-\frac{\spb{1}{Q^\flat}\langle{r}{(1+Q)}}
%{\bpa{1}{Q}{1}\spa{1}{r}}+\frac{m^2}{\spa{Q^\flat}{s}}
%\frac{\spa rs [1}{\bpa 1Q1 \spa 1r},\\
%\Jslash(Q^+;1,2)&=&-\frac{\spb{1}{Q^\flat}\langle{r}{(1+2+Q)}}
%{\bpa{1}{Q}{1}\spa{1}{2}\spa 2 r}+\frac{m^2}{\spa{Q^\flat}{s}}
%\frac{\spb 12\langle{s}{(1+2+Q)}}{\bpa 1 Q 1 \spa 1 2 ((1+2+Q)^2-m^2)}\\
%&-\frac{m^2}{\spa{Q^\flat}{s}}\frac{\spa rs [1}{\bpa 1 Q 1 \spa 12\spa 2r}\\
%\Jslash(Q^+;1,2,3)&=&\frac{\spb{Q^\flat}{1}\langle{r}{(1+2+3+Q)}}
%{\bpa 1Q1 \spa 12\spa 23\spa 3r}+\frac{m^2}{\spa{Q^\flat}{s}}
%\frac{\apb r21\langle{s}{(1+2+3+Q)}}{\bpa 1Q1\spa 12\spa 23
%\spa 3r((1+2+Q)^2-m^2)}\\
%&+\frac{m^2}{\spa{Q^\flat}{s}}\frac{\spa rs [1}{\bpa 1Q1\spa 12
%\spa 23\spa 3r}+\frac{m^2}{\spa{Q^\flat}{s}}\frac{\spa 2s\spb 21 [3}
%{\bpa 1Q1 \spa 12\spa 23((1+2+Q)^2-m^2)}\\
%&+\frac{m^2}{\spa{Q^\flat}{s}}\frac{[1Q(1+2)3]\langle{s}{(1+2+3+Q)}}
%{\bpa 1Q1 \spa 12 \spa 23((1+2+Q)^2-m^2)((1+2+3+Q)^2-m^2)}.
%\eeqa{massoffshell}
%Here, $1$, $2$ and $3$ are $+$ helicity gluons, $r$ is the reference 
%vector for the gluons and $s$ is the reference vector for the massive 
%fermion.  
However, rather than suggesting a general form for this current, the 
expressions seem to get only more complicated as the number of gluons 
increases.  It seems that for massive fermions, this method is not 
useful for determining the amplitude with all $+$ helicity gluons.

\Acknowledgements

The authors thank Jared Kaplan for helpful discussions of many 
parts of our formalism.  
A.~L.~thanks the University of Durham Institute for Particle Physics
Phenomenology for their tea and hospitality 
while some of this work was completed.
This work is supported by the US Department of Energy under 
contract DE--AC02--76SF00515.


\begin{thebibliography}{99}

%%
%%
\bibitem{AKR}
  D.~Atwood, A.~Kagan and T.~G.~Rizzo,
  %``Constraining Anomalous Top Quark Couplings At The Tevatron,''
  Phys.\ Rev.\  D {\bf 52}, 6264 (1995)
  [arXiv:hep-ph/9407408];
  %%CITATION = PHRVA,D52,6264;%%
 T.~G.~Rizzo,
  %``Searching for anomalous weak couplings of heavy flavors at the SLC and
  %LEP,''
  Phys.\ Rev.\  D {\bf 51}, 3811 (1995)
  [arXiv:hep-ph/9409460].
  %%CITATION = PHRVA,D51,3811;%%



\bibitem{HNW}
  P.~Haberl, O.~Nachtmann and A.~Wilch,
  %``Top Production In Hadron Hadron Collisions And Anomalous Top - Gluon
  %Couplings,''
  Phys.\ Rev.\  D {\bf 53}, 4875 (1996)
  [arXiv:hep-ph/9505409].
  %%CITATION = PHRVA,D53,4875;%%


\bibitem{BG}
  F.~A.~Berends and W.~T.~Giele,
  %``Recursive Calculations for Processes with n Gluons,''
  Nucl.\ Phys.\  B {\bf 306}, 759 (1988).
  %%CITATION = NUPHA,B306,759;%%

\bibitem{BCFW}
  R.~Britto, F.~Cachazo and B.~Feng,
  %``New Recursion Relations for Tree Amplitudes of Gluons,''
  Nucl.\ Phys.\  B {\bf 715}, 499 (2005)
  [arXiv:hep-th/0412308];
  %%CITATION = NUPHA,B715,499;%%
  R.~Britto, F.~Cachazo, B.~Feng and E.~Witten,
  %``Direct Proof Of Tree-Level Recursion Relation In Yang-Mills Theory,''
  Phys.\ Rev.\ Lett.\  {\bf 94}, 181602 (2005)
  [arXiv:hep-th/0501052].
  %%CITATION = PRLTA,94,181602;%%

\bibitem{Schwinn}
  C.~Schwinn and S.~Weinzierl,
  %``On-shell recursion relations for all Born QCD amplitudes,''
  JHEP {\bf 0704}, 072 (2007)
  [arXiv:hep-ph/0703021].
  %%CITATION = JHEPA,0704,072;%%


\bibitem{spinors}
  L.~J.~Dixon,
  %``Calculating scattering amplitudes efficiently,''
  arXiv:hep-ph/9601359;
  %%CITATION = HEP-PH/9601359;%%
  M.~L.~Mangano and S.~J.~Parke,
  %``Multi-Parton Amplitudes in Gauge Theories,''
  Phys.\ Rept.\  {\bf 200}, 301 (1991).

  \bibitem{infmom}
  J.~D.~Bjorken, J.~B.~Kogut, D.~E.~Soper,
  %``Quantum Electrodynamics at Infinite Momentum: 
%  Scattering from an External Field,''
  Phys.\ Rev.\  {\bf D3}, 1382 (1971);   J.~B.~Kogut, D.~E.~Soper,
  %``Quantum Electrodynamics in the Infinite Momentum Frame,''
  Phys.\ Rev.\  {\bf D1}, 2901-2913 (1970).


\bibitem{lightcone}
  S.~J.~Brodsky, H.~-C.~Pauli, S.~S.~Pinsky,
  %``Quantum chromodynamics and other field theories on the light cone,''
  Phys.\ Rept.\  {\bf 301}, 299-486 (1998).
  [hep-ph/9705477];   R.~Venugopalan,
  %``Introduction to light cone field theory and high-energy scattering,''  
  [nucl-th/9808023]; H.~Leutwyler,
  %``Mesons in terms of quarks on a null plane,''
  Nucl.\ Phys.\  {\bf B76}, 413-444 (1974).
  
\bibitem{SCET}
  C.~W.~Bauer, S.~Fleming, D.~Pirjol, I.~W.~Stewart,
  %``An Effective field theory for collinear and soft 
%gluons: Heavy to light decays,''
  Phys.\ Rev.\  {\bf D63}, 114020 (2001).
  [hep-ph/0011336].

  \bibitem{selfdual}
  W.~A.~Bardeen,
  %``Selfdual Yang-Mills theory, integrability and multiparton amplitudes,''
  Prog.\ Theor.\ Phys.\ Suppl.\  {\bf 123}, 1 (1996).
  %%CITATION = PTPSA,123,1;%%

\bibitem{explicit}
 To see explicitly that $\bar\sigma_{\mu\nu}F^{\mu\nu} = 0 $ in an 
external state of all positive helicity gluons, see \leqn{effhv}.
  
  \bibitem{cliff}
  C.~Cheung,
  %``On-Shell Recursion Relations for Generic Theories,''
  JHEP {\bf 1003}, 098 (2010).
  [arXiv:0808.0504 [hep-th]].
  
  \bibitem{cliffemail} C.~Cheung, private communication.
  
   \bibitem{CEK}   T.~Cohen, H.~Elvang, M.~Kiermaier,
  %``On-shell constructibility of tree amplitudes in general field theories,''
  [arXiv:1010.0257 [hep-th]].
  
  \bibitem{freddy}  P.~Benincasa and F.~Cachazo,
  %``Consistency Conditions on the S-Matrix of Massless Particles,''
  arXiv:0705.4305 [hep-th].
  %%CITATION = ARXIV:0705.4305;%%
  
  \bibitem{elvang} H. Elvang, ``On the structure of 
  scattering amplitudes'', talk given at Stanford University, January 25, 2010
  
 \bibitem{jared}
  N.~Arkani-Hamed and J.~Kaplan,
  %``On Tree Amplitudes in Gauge Theory and Gravity,''
  JHEP {\bf 0804}, 076 (2008)
  [arXiv:0801.2385 [hep-th]].
  %%CITATION = JHEPA,0804,076;%%

\bibitem{gravbcfw}
  J.~Bedford, A.~Brandhuber, B.~J.~Spence and G.~Travaglini,
  %``A recursion relation for gravity amplitudes,''
  Nucl.\ Phys.\  B {\bf 721}, 98 (2005)
  [arXiv:hep-th/0502146];
  %%CITATION = NUPHA,B721,98;%%  
  F.~Cachazo and P.~Svrcek,
  %``Tree level recursion relations in general relativity,''
  arXiv:hep-th/0502160;
  %%CITATION = HEP-TH/0502160;%%
  P.~Benincasa, C.~Boucher-Veronneau and F.~Cachazo,
  %``Taming tree amplitudes in general relativity,''
  JHEP {\bf 0711}, 057 (2007)
  [arXiv:hep-th/0702032].
  %%CITATION = JHEPA,0711,057;%%
  
  
\bibitem{perturbiner}
  A.~A.~Rosly and K.~G.~Selivanov,
  %``On amplitudes in self-dual sector of Yang-Mills theory,''
  Phys.\ Lett.\  B {\bf 399}, 135 (1997)
  [arXiv:hep-th/9611101].
  %%CITATION = PHLTA,B399,135;%%


\end{thebibliography}
\end{document}